# A Simulated Cyberattack on Twitter: Assessing Partisan Vulnerability to Spear Phishing and Disinformation ahead of the 2018 U.S. Midterm Elections

Michael Bossetta

# A Simulated Cyberattack on Twitter: Assessing Partisan Vulnerability to Spear Phishing and Disinformation ahead of the 2018 U.S. Midterm Elections

Michael Bossetta (University of Copenhagen)

**Abstract**
State-sponsored "bad actors" increasingly weaponize social media platforms to launch cyberattacks and disinformation campaigns during elections. Social media companies, due to their rapid growth and scale, struggle to prevent the weaponization of their platforms. This study conducts an automated spear phishing and disinformation campaign on Twitter ahead of the 2018 United States Midterm Elections. A fake news bot account – the @DCNewsReport – was created and programmed to automatically send customized tweets with a "breaking news" link to 138 Twitter users, before being restricted by Twitter.

Overall, one in five users clicked the link, which could have potentially led to the downloading of ransomware or the theft of private information. However, the link in this experiment was non-malicious and redirected users to a Google Forms survey. In predicting users' likelihood to click the link on Twitter, no statistically significant differences were observed between right-wing and left-wing partisans, or between Web users and mobile users. The findings signal that politically expressive Americans on Twitter, regardless of their party preferences or the devices they use to access the platform, are at risk of being spear phished on social media.

**Introduction**
After revelations of the Cambridge Analytica scandal and Russian-backed influence operations during the 2016 US election, social media platforms have increased their efforts to reduce the misuse of their platforms. Collectively, Facebook and Twitter have removed thousands of accounts linked to "bad actors,"[1] who engage in "platform manipulation"[2] to undermine trust in democracy. To date, much of the public's focus on bad actors has been on the paid use of trolls to spread propaganda (Aro, 2016; Zelenkauskaite and Niezgoda, 2018) or the abuse of platforms' advertising services by covert organizations (Nadler *et al.,* 2018).

However, bad actors also fashion social media into a much more concrete form of weaponry. State-sponsored cyber groups from Russia, Iran, and China increasingly weaponize social media platforms to conduct spear phishing attacks against Western governments (Bossetta, 2018). Spear phishing relies on social engineering – essentially a form of trickery – to bait victims into taking an action that reveals sensitive information. Automation is key feature of modern social engineering attacks, allowing attackers to conduct phishing attacks at scale (Ariu *et al.,* 2017).

Usually, phishing attacks occur through e-mail and rely on victims to click a malicious hyperlink, download an attachment laced with malware, or enter login credentials to a spoof website. If successful, a phishing attack can lead to the hijacking of a victim's social media, device, or private information.

Phishing remains the preferred method of state-sponsored actors to conduct cyberattacks. In 2017, "70 percent of successful security breaches associated with nation-state or state-affiliated actors involved phishing."[3] While difficult to quantify, only a small portion of these attacks likely occur via social media. Nevertheless, reports from cybersecurity firms estimate that spear phishing on social media rose 500 percent in 2016 (Proofpoint, 2017), tripled in 2017 (PhishLabs, 2018), dipped after platforms' purge of fake accounts, but increased 30 another percent in the first half of 2018 (Proofpoint, 2018).

Precisely due to platforms' efforts to remove bad actors and fake accounts, ordinary citizens on social media are now more valuable targets for state-sponsored phishing attacks than previously. The largescale removal of inauthentic accounts raises the currency of real accounts for bad actors. If bad actors want to spread disinformation without being detected, they can hijack real user accounts who have established an authentic history through their interactions with the platform over time.

Moreover, once a user's account has been successfully hijacked, bad actors can pivot off their success and launch successive attacks on that user's connections. Since users are more likely to open links from known connections rather than strangers (Seng *et al.*, 2018), bad actors can leverage compromised accounts to snowball an attack across a social network.

Taking seriously the political implications of large-scale cyberattacks on social media, the present study seeks to test the American public's vulnerability to spear phishing on Twitter. Therefore, I ask:

*How vulnerable are political Twitter users to spear phishing attacks on social media?*

To answers the research question, the study tests the extent to which partisan Twitter users are likely to click a hyperlink sent by a fake news account. The "DC News Report," an automated bot account created by the author, sent 138 Twitter users (77 right-wing partisans and 61 left-wing partisans) a link to a fabricated "breaking news" story about the 2018 Midterm Elections.

The results of the experiment reveal that 27 of the 138 users, or 20 percent, clicked the link. Three independent variables – partisanship, device, and time proximity to the election – were all found to be statistically insignificant predictors for clicking the link. This null finding suggests that the risk of being spear phished on Twitter cross-cuts partisan lines as well as the type of device used to access the platform.

Important to note is that the link was non-malicious and redirected users only to a Google survey form, which users were then invited to fill in. However, bad actors could easily circumvent the filters of link shortening services to weaponize the link by redirecting users to a malicious website that harbors a malware payload.

The study proceeds as follows. First, I outline the motivation for conducting a cyberattack experiment against the backdrop of researchers' increasingly limited access to social media data. Second, I outline why Twitter's digital architecture facilitates spear phishing attacks. Third, the experiment's methodology is outlined before the results are presented. Finally, the study concludes with a discussion of the study's findings.

**The case for experimental platform research**

Amidst the cross-platform crack down on bad actors and fake accounts, social media platforms have taken steps that limit researchers' access to data. Most pointedly, Facebook restricted access to its Pages API on 4 April 2018 (Schroepfer, 2018), ultimately barring researchers from accessing what little public data the platform offered previously.

Twitter, although maintaining a more generous data policy than Facebook, introduced a new verification process for app developers on 24 July 2018 (Roth and Johnson, 2018). While meant to curb the misuse of Twitter by malicious bot accounts, the initiative does little to help researchers help solve the automation problems that Twitter faces. Moreover, Twitter severely hampers researchers' access to data further back than one week in time, making it difficult to explain phenomena that are only brought to the public's attention after the fact (such as state-sponsored disinformation).

To be fair, both Facebook and Twitter have traditionally been somewhat generous in allowing researchers limited access to their data. However, the Cambridge Analytica scandal (which was in part caused by an academic [Wong *et al.*, 2018]) and the public outcry around state-sponsored disinformation have led social platforms to retreat further into their Walled Gardens.

Clearly, though, the scale of Facebook and Twitter has exceeded the companies' capabilities to competently monitor their own platforms. Time and time again, the platforms take steps to improve only after third-party actors, such as journalists, government agencies, and academics, alert the platforms to their own failures.

Facebook, for example, improved its advertising platform only after the investigative journalism organization ProPublica demonstrated that anti-Semitic categories could be targeted with ads on Facebook (Angwin *et al.*, 2017). Similarly, one week before the Midterm elections, Vice News exposed that the new "Paid for by" feature required by Facebook to run political ads – an effort aimed to improve transparency (Leathern, 2018) – was flawed. Vice News successfully ran ads identical to those issued by the Russian Internet Agency and attributed them to being "paid for by" prominent US Senators (Turton, 2018).

Twitter, although less a target for investigative journalism than Facebook, has unknowingly hosted a Russian-led cyberattack targeted at over 10,000 Pentagon employees (Calabresi, 2017). And even though the platform is active behind the scenes in shutting down bots, the scholarly work of Bastos and Mercea (2017) was integral to alerting British policymakers of partisan botnets active during the Brexit referendum.

The point here is not to criticize the performance of Facebook and Twitter. Rather, my aim is to highlight that these platforms rely on external help to detect problems on their platforms and refining their policies. The magnitude of data flowing through tech giants like Facebook and Twitter are unprecedented in history, and they seemingly exceed the monitoring capabilities of any one company or service.

Look no further than recent initiatives by Facebook and Twitter to enlist the help of select researchers to solve some of the platforms' most pressing problems – in exchange for money and privileged access to data. Facebook is the first partner in the recently established Social Science One initiative, which aims to assist platforms to "produce social good, while protecting their competitive [market] advantage."[4] The initiative is focused specifically around social media's impact on elections and democracy.

Twitter, meanwhile, has recruited two teams of researchers from 230 proposals to "more deeply understand the concept of measuring conversational health,"[5] likely in response to growing public controversies around free speech censorship on the platform.

To summarize the above, Facebook and Twitter have failed at safeguarding the integrity of their platforms for users. Historically, these platforms have relied on external watchdogs to point out flaws in their systems to improve their policies. As the platforms throttle researchers' public access to data, Facebook and Twitter's new incentive structure is to hire academics and grant them privileged access to data.

These steps may be positive for the platforms in the long run, but state-sponsored bad actors do not play by the same rules in the short term. As select researchers slowly obtain and analyze data provided by Facebook and Twitter, bad actors are running experiments in real time with the purpose of influencing elections and undermining trust in democracy.

Therefore, I advocate meeting bad actors where they are and running ethical, non-manipulative experiments on social media during contemporary political events for two reasons. First and on the supply side, this method tests the responsiveness of platforms in shutting down malicious actors (which, as will be shown in this experiment, Twitter was relatively quick to do). Second and on the demand side, live experiments conducted on social media afford researchers the possibilities to understand, test, and explain mechanisms that might help safeguard the public against bad actors in the future.

This study, employing a relatively basic and small-scale experimental design in comparison to state-sponsored actors, tests the vulnerability of partisan Twitter users to click malicious links from a fabricated news outlet. The study is a live field experiment that simulates a spear phishing cyberattack, and in order to accurately assess citizens' vulnerability, the users selected for the study were not notified before the experiment.

However, users' anonymity is protected and no personally identifiable information is reported in the study. The project received ethics approval from the University of Copenhagen and was conducted in full compliance with the European General Data Protection Regulation (GDPR).

Conducting live field experiments on social media platforms is not unprecedented. In cybersecurity studies, researchers have simulated phishing attacks on Twitter (Seymour and Tulley, 2016) and Facebook (Benenson et al., 2014) without obtaining prior consent from users. Similarly, in political communication studies, Vaccari *et al.* (2017) and Chadwick *et al.* (2018) solicited 22,000 and 39,639 Twitter users to participate in a survey via several automated Twitter accounts.

The ability to automatically initiate contact with thousands of users on a non-paid basis is a particular feature of Twitter's digital architecture. In the following section, I outline Twitter's architectural features and explain why the platform is prime for spear phishing by bad actors.

**Twitter: The perfect digital architecture for automated spear phishing**

Despite the glitzy features that users engage with on the front end, social media platforms are constructed from thousands of lines of unglamorous computer code. I refer to this back end code as a platform's digital architecture: "the technical protocols that enable, constrain, and shape user behavior in a virtual space."[6]

Every social media platform has a distinct digital architecture, which is constantly improved to achieve or maintain a competitive market advantage. A platform's digital architecture directly impacts its available features, what they do, and how they work. As such, the digital architecture of a platform sets the parameters for how users create, engage with, and distribute political content on both the Web and mobile version of a social networking site (Bossetta, Dutceac Segesten, and Trenz, 2017).

Typically, the average social media user engages with the features of a platform on the front end through its *Graphical User Interface* (GUI). For example, a user may signal a reaction to a Facebook post in the News Feed by tapping the "like" button on a mobile phone, or share a tweet to one's followers on Twitter by clicking "retweet" with a mouse connected to a desktop.

However, most platforms allow for users to interact with the platform on the back end through an *Application Programming Interface* (API). APIs allow users with programming knowledge to collect large amounts of data from the platform in a much faster manner than copying and pasting information from the GUI. In addition, most APIs also permit users to control an account directly through the platform's back end by writing computer code, without having to access the GUI.

The ability to control an account content through an API opens up the possibility for account automation. On Twitter, for example, a user can write computer code for an account to automatically post, favorite, or retweet tweets, as well as send direct messages to other users. An automated account is typically referred to as a bot.

As with a platform's front end, the digital architecture of a platform also sets the criteria for APIs and subsequently, automation. Relative to other social media platforms, Twitter's digital architecture permits a high degree of automation and therefore hosts a "bot-friendly API".[7] As a result, Twitter hosts millions of bots as part of its 326 monthly active users (Twitter, 2018), although the exact number of bots on the platform is disputed.

Before Twitter's bot crackdown, Varol *et al.* (2017) estimated that bots comprised "between 9 and 15 percent of active Twitter accounts are bots."[8] Twitter, in a Senate hearing held on 29 November 2017 regarding social media's influence in the 2016 US election, rebutted this claim and stated that less than 5 percent of its monthly active user were bots (Edgett, 2018).

While programmers write bots on Twitter for a wide array of purposes, such as comedic entertainment or marketing, bots have also been deployed for a wide array of political purposes such as spreading disinformation or propaganda (for an overview, see Woolley, 2016). Political bots have been found to push an agenda in online debates leading up to elections (Bessi and Ferrara, 2016) as well as non-political contexts such as discussions around fandom (Bay, 2018).

Apart from electoral influence, automated bot accounts can be weaponized to conduct spear phishing attacks. Chhabra *et al.* (2011) refer to Twitter as a "phisher's paradise,"[9] finding in their study that 89% of the accounts engaging in phishing on Twitter were automated. Apart from automation, five other aspects of Twitter's digital architecture make the platform technology conducive for spear phishing.

First, Twitter's APIs allow for the collection of structured data on its users, which can be used to both discover potential targets for a spear phishing attack as well as perform reconnaissance on their whereabouts, interests, and connections (Bossetta, 2018). The data mining of information about users, a technique referred to as Open Source Intelligence (OSINT) is a common tactic in modern phishing attacks (Ariu *et al.*, 2017).

Second, Twitter's digital architecture supports hyperlinking through short URLs such as those generated by https://bitly.com. Phishers utilize short URLs to obfuscate the identity of a malicious link (Nepali and Wang, 2016), and due to Twitter's 280 character limit, short URLs are commonly used on the platform.

Third, phishers can target a tweet with a malicious short URL to a particular user directly with the @mention feature, which sends a notification to that user and increases the change the targeted user will see the tweet. Moreover, if a tweet begins with an @mention, the tweet "is visible in most circumstances only to the sender and addressee,"[10] essentially hiding the attack from public view.

Fourth, users' privacy settings on Twitter are set to open by default. Phishers can therefore @mention a targeted user without establishing prior contact.

Fifth, Twitter's 280 character limit, as well as its timeline feed's focus on chronology, create a platform profile where users go for breaking news (Osborne and Dredze, 2014). As a result, Twitter attracts a user base who spend time on the platform to satisfy a need for cognition (Hughes *et al.*, 2012). Since users regularly log into the platform for news and fulfill a cognitive need, they are primed to click links that seem to promise access to information.

Altogether, six elements of Twitter's digital architecture support spear phishing on the platform: account automation, open source intelligence gathering, short URL support, @mention notifications, open network structure, and a news-oriented content profile. In the following section, I outline how each of these architectural features were leveraged for the simulated cyberattack in this study.

**Methodology**

*Constructing the Cyberattack*
This study's methodology is heavily inspired by the work of cybersecurity researchers Seymour and Tulley (2016). They developed SNAP_R, a machine learning based Python tool that "automatically generat[es] spear-phishing posts on social media."[11]

Succinctly put, SNAP_R operates as follows. A list of Twitter accounts is fed into the tool, which connects to Twitter's REST API and collects the recent tweets of each account. Then, the tool uses the Markovify Python module (Singer-Vine, 2015) to generate random tweets based an algorithm that combines words and sentences that the user has expressed in previous tweets. As a result, the generated tweets mimic the language and interests expressed recently by the user, but they are varied enough as to not replicate any previous tweet. No output from the generated tweets is reported here to protect users' anonymity.

Once tweets have been generated for each user, SNAP_R adds an @mention to the beginning of the tweet and appends a short URL to the end of the tweet. I reprogrammed the tool to also add an emoji of a hand pointing downwards (👇), to draw users' attention to the tweet and the chances of clicking. The resulting tweet can be summarized as follows:

@TargetUser + Tweet generated by Markovify + 👇 + Short URL

SNAP_R then leverages Twitter's automation feature to send tweets to each user. Since an @mention begins each tweet, the user is alerted via a notification that they have been contacted. Moreover, and as mentioned in the previous section, tweets beginning with an @mention are largely hidden from public view.

The tweets generated and issued by SNAP_R are not directly visible on a users' timeline; rather, they are located in the "Tweets & Replies" section of the graphical user interface. This means that if you were to visit the Twitter profile of user who has been sent a tweet via SNAP_R, you would have to actively navigate to the "Tweets & Replies" section of that user's profile in order to know that the tweet existed.

Another important feature to note relates to the tool's short URL creation. SNAP_R uses the Google's link shortening service, [https://goo.gl](https://goo.gl), which has three implications for spear phishing. First, the service hides the link's real destination. Second, the real link's destination hides behind a well-known brand (i.e., Google). Third, the goo.gl service allows the tweet's sender to monitor information such as when the link was clicked, from where, as well as from what type of device. This information can be used by spear phishers to detect patterns in clicks and optimize future attacks.

For this project, though, SNAP_R needed to be reprogrammed since Google discontinued its goo.gl link shortening service and instituted a new service called Firebase Dynamic Links.[12] Firebase Dynamic Links (FDL) offer much less detailed analytics than the goo.gl service; FDL only allows users to see the overall number of clicks received by a link.

However, FDL has the benefit of allowing users to customize exactly how short links are displayed on social media. Users can define the exact picture, headline, and byline of how a link will display on Twitter. Since this study's interest concerns Twitter users' likelihood to click information from a fake news outlet (discussed below), I customized the FDL to display on Twitter as shown in Figure 1.

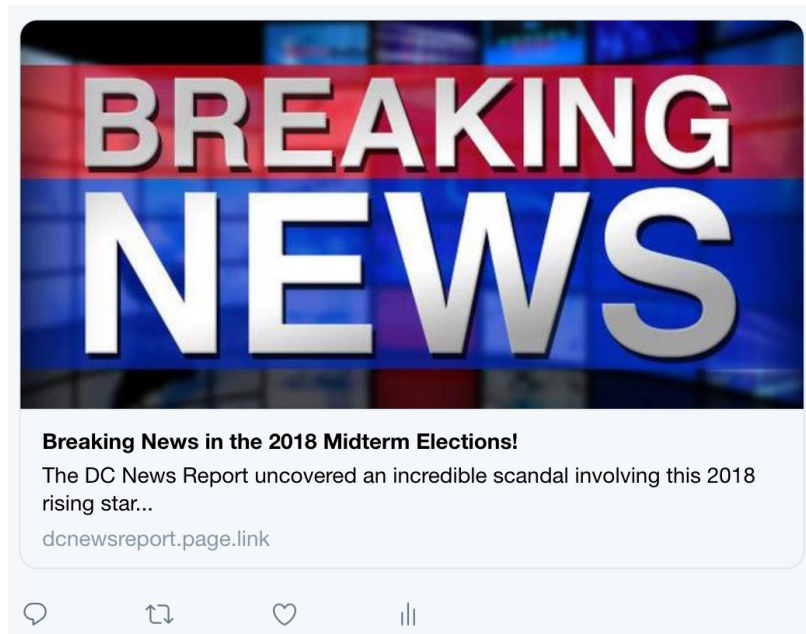

Figure 1: Firebase Dynamic Link as shown on Twitter

The embedded link's picture and text were designed to convey political neutrality. Although no pre-testing was performed prior to the experiment to verify the link's neutrality, the following steps were taken to minimize harm while maintaining ecological validity.

A scandal involving a "rising star" was chosen as the text for the byline, as such a news item would believably warrant the status of "breaking news" in today's sensationalist media landscape. While the byline is misleading in referring to a non-existent scandal, harm is minimized by not mentioning any candidate or party involved in the scandal, so as to not influence a user's political opinion.

Similarly, the "breaking news" image contains the colors of both major political parties (red and blue) and offers no further contextual information as the background is blurred. The image was selected from a Google Images search for "breaking news." Although its copyright status could not be verified, the image appeared on multiple private blogs and is consided by the author to fall under copyright fair use for nonprofit, educational purposes.

Important to emphasize is that the link did not lead to any malicious destination; rather, it directed users to a short Google form that required answering three questions. The first asked users about the device they used to access the link: computer, smartphone, or tablet. The remaining two questions focused on users' social media activity. One question asked users how often they checked Twitter, and the other asked users to check with other social media platforms they use. Figure 2 below displays the Google Form.

## Thanks for checking out the DC News Report!

You've been selected by a Twitter Bot to participate in a study conducted by the University of Copenhagen.

We're asking both liberals and conservatives to answer three quick questions about how they use social media. All responses will be anonymized, and if you do not wish to participate, simply close this tab. Thanks!

* Required

**What device are you on right now?** *

○ Smartphone
○ Computer
○ Tablet

**How often do you check Twitter?** *

|  | 1 | 2 | 3 | 4 | 5 |  |
|---|---|---|---|---|---|---|
| Less than once per Week | ○ | ○ | ○ | ○ | ○ | More than once per Day |

**Which social media platforms do you also use?** *

☐ Facebook
☐ Instagram
☐ Reddit
☐ LinkedIn
☐ Snapchat
☐ Other: ______________

**Thank you for your time! Feel free to leave us any feedback :)**

Your answer

[SUBMIT]

Never submit passwords through Google Forms.

**Figure 2: Google Form survey**

*Account Creation and Mock-Up*

The following section details the process of creating the @DCNewsReport Twitter account. The aim was to create an account using (mostly) free and anonymous methods, in order to make tracing the account to any individual person or entity difficult.

All Twitter accounts must be verified by providing a valid email address. For this purpose, a Google Gmail account was created, which in turn requires verification by mobile text message. To verify the Gmail account, the website [https://receivesms.xyz](https://receivesms.xyz) was used to receive the verification code to an online number associated with the site. Several of these websites exist and are often used to verify social media accounts without using one's personal mobile number. Once the Gmail account was verified and activated, the @DCNewsReport Twitter account could be verified through email.

With the @DCNewsReport up and running, the next step was to mimic a legitimate US news organization. After browsing several American media outlets on Twitter, The Washington Post (@WashingtonPost) was chosen as a template due to its easily reproducible profile picture and cover photo, which featured the United States Capital building in Washington. A freelance graphic designer, contracted via the website Fiverr.com, created a similar profile picture and cover photo for $5.

Figure 3 below depicts the real @WashingtonPost account and the created @DCNewsReport (all identifying information to real Twitter accounts in the photo have been blurred). Important to note is that the @DCNewsReport did not contain any profile description information, in order to avoid impersonating an authentic journalism outlet.

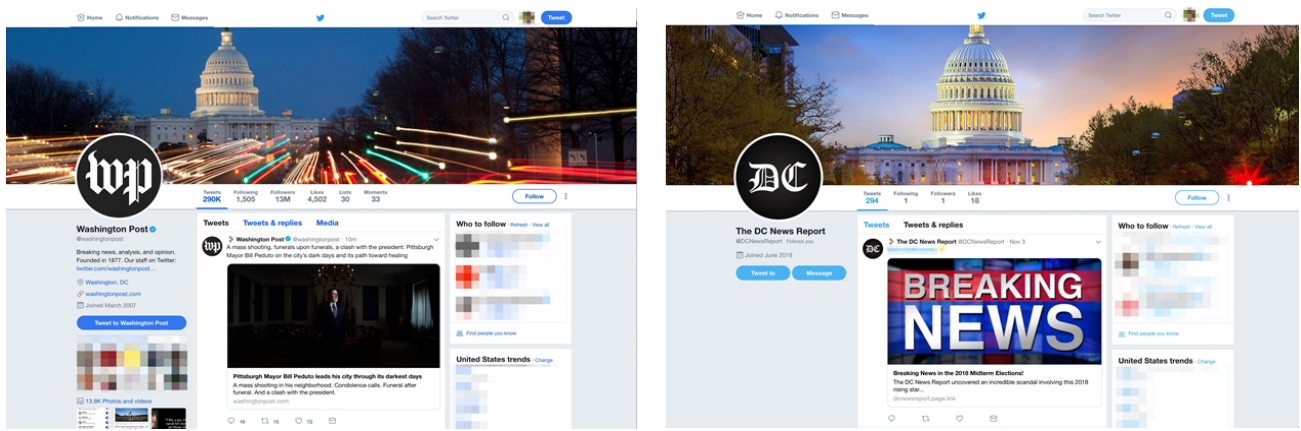

Figure 3: @WashingtonPost and @DCNewsReport Twitter Accounts

In order to post Twitter messages automatically through the platform's API, one needs to also register a Twitter app developer account. As of July 2018, and likely due to the bot crackdown mentioned earlier, app developers must apply through Twitter and state their motivations for creating a developer account. However, the DCNewsReport was established in June 2018 and only needed to undergo another mobile verification process to create an app developer account on Twitter.

This time, attempts to use online websites to verify the number failed. Instead, Google Voice was used to generate a number that successfully validated the DCNewsReport's Twitter developer account.

Apart from posting messages automatically, a Twitter developer account allows you to collect data at scale through Twitter's freely accessible APIs. The Streaming API collects data in real time based on certain keywords. The Search API, meanwhile, collects data 7-10 days in the past. However, neither API delivers all tweets based on a search criteria. For a full list of tweets, you need to pay for access to the Firehose API, which can cost thousands of US dollars.

Thus, for this project, only the Streaming API and Search API were used for data collection. Although only a subset of overall tweets containing the search criteria were collected, the aim was only to collect a sample of Twitter users tweeting about the 2018 US Elections. The data is collected over two time periods and divided into two studies. Although the methodologies of both studies are similar, the first study was run on a small number of users as a pilot to avoid bot detection. As will be shown by the second study, this concern was warranted as the @DCNewsReport was eventually blocked from posting tweets automatically.

*Study 1:*

*Data Collection and User Selection*
For Study 1, Twitter's Streaming API was queried using the package "rtweet" (Kearney, 2018) for the programming software R. The search query focused on tweets containing the word "midterm" (and the plural "midterms") from 18 September 2018 – 26 September 2018. This time period was heavily focused on controversy around Brett Kavanaugh, then a Supreme Court nominee fighting accusations of sexual harassment. The dataset originally included 393,872 tweets (and retweets) from 182,541 unique users.

Twitter would likely detect an account issuing unsolicited tweets (i.e. "spam") to such a large a number of users. Therefore, several steps were applied to filter the data. First, retweets were removed, as well as duplicate tweets. This resulted in 59,913 tweets from 46,132 users.

To further divide the data, only accounts having more than 50 followers and less than 400 were kept. This step was taken for two reasons. First, removing accounts with over 50 followers helps reduce the number of bots accounts in the dataset. Many bots – including the one developed for this study – are not aimed at attracting followers; they simply spam messages to other users. On the other hand, keeping only users with less than 400 followers helps ensure that organizations, celebrities, and brands are not included in the data. Therefore, the resulting 15,153 users are likely to be real, ordinary Twitter users.

The timelines of each of these remaining users were harvested, and only tweets sent after June 2018 were retained. From this subset, users were kept in the dataset if they tweeted "MAGA" or "BlueWave", offering a preliminary signal as to whether they were right-wing or left-wing users, respectively. MAGA stands for "Make America Great Again" and is often used a slogan by Trump Supporters. On the other side of the aisle, "BlueWave" is a reference to Democratic Party supporters voting in the elections to take power back from Republicans. As a result of these filtering steps, the final tally of MAGA tweeters was 113, and 120 Blue Wave tweeters.

The next step of the method aimed at selecting a small number of users to test the study's hypotheses while avoiding bot detection. A manual coding was performed in order to make sure users met three criteria: they often tweeted about politics, their ideological views were consistent, and they tweeted from the same device.

For each of the MAGA and BlueWave tweeters, a random sample of ten tweets were taken and compiled into an Excel spreadsheet. To be included in the final dataset, 80 percent of the sampled tweets for each user needed to be about politics, express a political preference (Republican or Democrat), and be sent from the same device. 51 MAGA tweeters and 31 BlueWave tweeters met this criteria, mostly since the BlueWave tweeters issued their tweets from a diverse array of devices (e.g. the Web, smartphones, and iPads).

Right-wing users typically expressed tweets along issues such as: support for Donald Trump, the confirmation of Brett Kavanaugh, protest of Nike for supporting Colin Kaepernick, the use of the MAGA hashtag, criticism of migrant caravans, and conservative codewords like "snowflake" or "social justice warrior."

Left wing users, meanwhile, could be identified through tweeting expressions that signaled: criticism of Donald Trump, the rejection of Brett Kavanaugh, support for environmental protection legislation, support for the #metoo movement, and calls to action for Democrats to vote in the midterms (i.e. create a "blue wave").

These remaining partisan users who tweeted from the same device were randomly sampled into four batches of ten users. This step was taken so that each batch of users would be "attacked" on a weekday (Monday – Thursday). Every batch had five web tweeters (tweeting from a computer) and five mobile tweeters (tweeting from an Android or iPhone). To segment by ideology, two of the batches included right-wing partisans while the other two comprised left-wing partisans. The final breakdown of users can be summarized as follows:

$$40 \text{ Users} = (5 \text{ Web} + 5 \text{ Mobile}) \times 2 \text{ Ideology} \times 2 \text{ Days}$$

Lastly, each user was run through the [Botometer bot checker](#) (Varol *et al.,* 2017) to ensure that users were authentic. Botometer gives accounts a score from zero to five, with accounts closer towards five indicating bot-like activity. The mean score for the 40 users was .66, indicating most accounts were authentic. Only one account scored over two (receiving a 3.6), but upon a manual inspection it was unclear whether the account was a bot or not. Thus, that user was left in the study.

### *Launching the Attack*
The first study ran between 22 October – 24 October 2018 (Monday – Thursday). Right-wing users were targeted Monday and Wednesday, and left-wing users were targeted Tuesday and Thursday. Every day, the first tweet was sent at 3pm EST. This time was chosen because according to an analysis by the social media management platform Hootsuite, "The best time to post on Twitter [for engagement] is 3pm Monday to Friday."[13]

Previous research suggests less than one percent of Twitter users tag their tweets with geo-location information (Cheng *et al.* 2010). Therefore, no attempt was made to segment users by time zone. Instead, the attack began at 3pm EST and as discussed below, would continue for a time period corresponding to random sleep intervals. Thus, users living west of EST might have been targeted around 3pm in their corresponding time zones as the attack progressed, but there is no simple way to discern when users actually came into contact with the tweet.

Once the first tweet was issued, the bot would "sleep" (i.e., remain inactive) for a random interval between 61 and 950 seconds in an attempt to avoid bot detection by Twitter. Similarly, after each attack tweet generated by SNAP_R, a tweet without an @mention or link would be issued at the same random interval. These tweets would appear on the @DCNewsReports main timeline and were randomly selected from a list of 100 tweets created by the author in a .txt file.

The tweets in the .txt file had small variations in content such as: "The DC News Report checks our facts. We work around the clock to bring you breaking news!" and "We check our facts. That's why we're proud to be the DC News Report!". The purpose of these text-only tweets was to avoid Twitter's bot filters, which actively seek out accounts who consistently @mention users or spam links.

Each batch of ten users received its own Firebase Dynamic Link to separate any survey responses into batches. Due to how Twitter embeds links in tweets, the hyperlink itself was not shown to the user. In text form, the user only saw the @mention, the generated tweet, and the hand pointing emoji. The FDL was then embedded into the picture shown above in Figure 1.

Link clicks were measured using Twitter's own "Tweet activity dashboard."[14] The feature allows a tweet sender to see how many times a link in a tweet was clicked, as well as other information such as number of retweets, likes, and profile clicks. Any engagement with a tweet listed by the dashboard was recorded except for impressions, which were influenced heavily by the researcher's own monitoring of the tweets.

### *Results (Study 1)*

Out of the original 40 users, 39 successfully received the tweet. One user's account was protected (i.e., private and requiring manual approval to receive tweets), and therefore was unable to be targeted.
Overall, nine out of 39 users, or 23 percent, clicked the link included in the targeted tweet. Six were right-wing partisans, and three were left-wing partisans. Looking at device type, five users who clicked the link were desktop users, and the other four were mobile phone users. In terms of other engagements, only two users out of the 40 sought more information about the tweet's source by clicking on the @DCNewsReport's profile.

Interestingly, four users from the first batch of right-wingers filled out the Google Forms survey, but no other users did. The survey respondents all indicated they check Twitter "multiple times per day" and are also active on Facebook and LinkedIn. Two indicated being active on Instagram, and one user reported being active on Reddit.

While these results offer some preliminary insights, a trial of 39 users is insufficient to make any generalizations. The main purpose was to see whether such an experiment could be run while avoiding Twitter's bot detection. Overall, the @DCNewsReport generated 78 tweets (39 attack tweets with links, and 39 tweets with text-only tweets from the .txt file). Therefore, a second study was conducted in order to increase the number of participants, as well as uncover whether tweets sent in closer proximity to the midterms would increase the click rate.

*Study 2*

Study 2 used the same search criterion as Study 1 (keyword: "midterm") but this time used the Search API to collect tweets from 28 October 2018. Time constraints necessitated collecting data from the Search API, which collects data in the past. Once bot detection was successfully avoided in the first study, a larger scale attack aimed to increase the number of targeted users in the final week before the midterm.

Thus, to perform the second study before the midterm date of 6 November 2018, data needed to be collected historically. 145,818 original tweets (i.e., no retweets) from 94,036 users were collected, since the Search API allows you to filter out retweets when collecting data. After applying the same filtering steps from Study 1, Study 2 resulted in 277 MAGA tweeters and 153 Blue Wave tweeters.

Like in Study 1, these users' timelines were harvested and filtered to include only tweets from 1 September 2018 to be recent. A random sample of 10 tweets per user were loaded into an Excel file, and users were coded by partisanship and device type. In both cases, at least eight tweets needed to signal a clear partisan affiliation of tweet publication from the same device.

The coding resulted in 198 users. 30 right-wing and 30 left-wing partisans tweeted solely from a desktop, whereas 86 right-wing users and 52 left-wing users could be classified as mobile users. In Study 1, some mobile users were discarded from the study in order to keep the normalize the distribution of left-and right-wing, as well as desktop and mobile. In order to increase the participants in Study 2, however, all mobile users were included, leading to a mobile user population that skewed right-wing.

The 196 users were again divided into four batches over four days:

196 Users = (15 Web + ½ of Mobile per Ideology) x 2 Ideology x 2 Days

However, Twitter restricted the @DCNewsReport from posting automatically on the third day of Study 2, and therefore only the first two days are reported. The breakdown of the users from the first two days (n=98) are reported in Table 1.

| Ideology | Desktop | Mobile |
|---|---|---|
| Right-wing | 15 | 42 |
| Left-wing | 15 | 26 |

Table 1: Study 2 participants by ideology and device

Study 2 ran on 29 and 30 October 2018, with the first tweet issued at 3pm EST on both days. The average Botornot score for the users in Study 2 was .64, with 1.7 as the highest score. Thus, the probability is high that none of the users included in this study were automated accounts.

*Results (Study 2)*

In Study 2, 18 out of 98 users clicked the link, or 18 percent. 10 were right-wing partisans, and eight were left-wing partisans. Given that more right-wing partisans (n=57) were targeted in Study 2 than left-wing partisans (n=41) in order to increase the sample size, the difference between the two ideologies' likelihood to click becomes rather marginal at 2 percent (17 percent or right-wingers, 19 percent for left-wingers). Looking at device type, six of the 30 desktop users clicked (20 percent), compared to 12 of the 68 mobile users (17 percent). Irrespective of ideology or device type, approximately one in five users were tricked into clicking the link. Only three right-wing and two-left wing partisans sought more information about the @DCNewsReport by clicking the account's profile.

Interestingly, one left-wing user both liked and retweeted the link, but that user (nor any of that users' followers) appeared to click the link. Moreover, one right-wing user who clicked the link, and also checked the @DCNewsReport's profile three times, replied to the link. The reply angrily called out the account for being fake, associated the account with Democratic party, included profanity, and used a racial slur toward Hispanics. The generated tweet issued with the link related to the migrant caravan, and the ethical implications of the using machine learning in live experiments on social media is discussed in the final section.

Like in Study 1, four users filled out the Google Forms survey but this time from the left-wing batch. All four participants expressed that they checked Twitter "multiple times per day." Three reported also using Facebook, two reported using Instagram and LinkedIn, and one reported using Reddit. Below, I aggregate the results of the two studies to paint an overall picture of the findings.

*Results (Overall and statistical tests)*

In total, 27 out of 138 users (or 19.5 percent) clicked the link. Table 2 reports the overall Click Through Rate (CTR) for each group of targeted users, divided by: Ideology, Device Type, and Study (a proxy for time relative to election day). The CTR calculates the percentage of users who clicked in relation to the overall number of users in each group.

| Study | Ideology | Desktop Users | | | Mobile Users | | |
|---|---|---|---|---|---|---|---|
| | | Click | Total | CTR | Click | Total | CTR |
| 1 | Right | 3 | 10 | 30% | 3 | 9 | 33% |
| 1 | Left | 2 | 10 | 20% | 1 | 10 | 10% |
| 2 | Right | 4 | 15 | 27% | 6 | 42 | 14% |
| 2 | Left | 2 | 15 | 13% | 6 | 26 | 23% |

Table 2: Overall results and Click Through Rate (CTR) by Ideology, Device, and Study

Although the CTR's differ across time, ideology, and device, the relatively small number of users included here do not allow for making sweeping generalization about user behavior. Twitter ultimately prohibited more users from being included in the study, as the platform restricted the @DCNewsReport from issuing automatic tweets on the seventh day of the experiment.

To examine whether ideology, device type, or time proximity to the election could explain users' likelihood to click the breaking news link, chi-square tests were performed on the overall results and each variable separately to see whether any were statistically independent. Ideology, device type, and time were coded as binary variables and tested for independence against click status. The results of the chi-square tests are reported in Tables 3, 4, and 5.

*Results of Chi-square Test and Descriptive Statistics for Click by Ideology*

| Click Status | Ideology | |
| --- | --- | --- |
| | Right-Wing | Left-Wing |
| Click | 16 (21%) | 11 (18%) |
| No Click | 61 (79%) | 50 (82%) |

*Note*. $\chi^2$ = 0.163, df = 1, p = .686. Numbers in parentheses indicate column percentages.
*p < .05

Table 3: Results of chi-square test for click by ideology

*Results of Chi-square Test and Descriptive Statistics for Click by Device*

| Click Status | Device | |
| --- | --- | --- |
| | Desktop | Mobile |
| Click | 11 (23%) | 16 (18%) |
| No Click | 37 (77%) | 74 (82%) |

*Note*. $\chi^2$ = 0.525, df = 1, p = .469 Numbers in parentheses indicate column percentages.
*p < .05

Table 4: Results of chi-square test for click by device

*Results of Chi-square Test and Descriptive Statistics for Click by Time*

| Click Status | Time Period | |
| --- | --- | --- |
| | Study 1 | Study 2 |
| Click | 9 (23%) | 18 (20%) |
| No Click | 30 (77%) | 81 (80%) |

*Note*. $\chi^2$ = 0.426, df = 1, p = .541 Numbers in parentheses indicate column percentages.
*p < .05

Table 4: Results of chi-square test for click by time period

The chi-square tests show that there were no significant differences between users' *ideology* ($X^2$ (1, N = 138) = 0.163, p > .05), their *device type* ($X^2$ (1, N = 138) = 0.525, p > .05), or the *time proximity* to the election day ($X^2$ (1, N = 138) = 0.426, p > .05). While minor differences can be observed in the percentage of users' who clicking the link based on the variables, the percentages ranged only between 18-23 percent. This null finding, and its relevance, is interpreted in the following and final section.

*Discussion and Conclusion*

Combining the results from the two studies, 138 Twitter users were automatically issued personalized tweets that contained a link to "breaking news" about the 2018 US Midterm Elections. The aim of the experiment was to answer the research question: *How vulnerable are political Twitter users to spear phishing attacks on social media?*

Overall, 27 out of 138 users (or 19.5 percent) clicked the link that could, if issued by a bad actor, have contained a malware payload. The results of this experiment therefore suggest that *one in five* political Twitter users are susceptible to spear phishing attacks.

Moreover, there were no statistically significant differences found between users' likelihood to click the link based on ideology (right-wing or left-wing), device type (desktop or mobile), or time proximity to the election day (one or two weeks before the vote). This null finding suggests that political users on Twitter are equally susceptible to spear phishing attacks, regardless of their ideology of the device used to access the platform.

This finding is surprising, as most studies argue that right-wing users are more likely to share disinformation (Badaway *et al.,* 2018; Narayan *et al.,* 2018). Thus, we might expect that right-wing users are also more likely to click disinformation. However, the study's design and results point to an important difference between the sharing and actual clicking of disinformation content.

Studies of disinformation tend to find that right-wing Twitter users are more likely to retweet - and thus amplify - dubious information. However, they do not offer a sense of how likely users are to *click* such stories. One study on Twitter estimates that 59% of URLs on Twitter are never clicked (Gabielkov et al., 2016), suggesting that sharing information on Twitter is not necessarily correlated to reading it (the study even found one example for this in the case of a left-wing user who retweeted the link without clicking it). Although previous research suggests that right-wing users might be more apt to share disinformation on Twitter, the findings reported here demonstrate that users across party lines are susceptible to clicking a disinformation news item.

In addition, users of both political persuasions were highly unlikely to validate the source of disinformation. Only 5 percent of users clicked on the @DCNewsReport's profile to check the account's authenticity. Had users visited the account's profile, they would have quickly uncovered that the outlet was not legitimate, since it only had one follower (the author) and contained no serious information.

Also worthy of note is that eight of the 27 users who clicked the link (or 30 percent) filled out the survey. While their responses varied on two of three questions, the survey participants were unanimous in stating that they checked Twitter "multiple times per day." This result appears to align with the results of a study by Vishwanath (2014), who showed in an experimental design that habitual Facebook use correlated positively with the likelihood be spear phishing on the platform. It was not feasible to survey those who didn't click this study's fake news link, but future research should examine more thoroughly how the social media habits of individuals relate to their susceptibility to spear phishing and disinformation on social media.

Although exploratory and involving a small number of participants, the current study paints a bleak picture for Twitter users' susceptibility to cyberattacks and attentiveness to disinformation. Though only 20 percent of users clicked the link, their accounts or devices could have been opened to compromise by bad actors, which has two serious implications.

First, bad actors could leverage an account or device takeover to inject ransomware or steal sensitive information, potentially leading to identity theft. Second, the connections of a compromised account may also be put at risk, with bad actors using the account to target other users in a network. In the context of election interference, compromised accounts could be manipulated to spew propaganda or disinformation. Since users are more likely to be mobilized in the democratic process when campaign messages are mediated through social media connections (Aldrich *et al.*, 2016), disinformation spread by a users' connections might be more persuasive than the same content issued through strangers. Moreover, hijacked accounts with a history of engagement with platform would be extremely difficult for Twitter's filters to detect, allowing the hijacked accounts to operate for an extended period of time.

Adding to these concerns, the cyberattack simulated here was extremely basic and low-threat by design. State-sponsored bad actors with a more resources, technological proficiency, and malicious intent could conduct a much more sophisticated attack involving multiple bot accounts and higher level targets, such as government personnel. Indeed, previous media and cybersecurity reports signal that such attacks targeting the Pentagon have already occurred on Twitter (Calabresi, 2017; ZeroFOX, 2017).

This study replicated a technique likely used in these Russian-backed attacks on the US government employees: the automatic generation of tweets using machine learning. Given the relatively black-box mechanics of machine learning, the use of such tools for research raises critical questions about ethics. For example, although this study sought to minimize harm and neutralize partisanship in the link's stimuli, the automated text generation clearly angered one user, who replied with profanity to the @DCNewsReport's tweet. While the generated tweet derived solely from that user's previous language, the output of the machine learning algorithm indirectly framed President Trump in a denigrating way that the user had previously expressed about someone else. Yet, given the study's focus on selecting political partisans, the issued tweet was unlikely to change the user's political preferences.

The digital architectures of social media, and the manipulation of them, afford researchers the possibility to conduct live experiments on platforms. In some cases, this is necessary to answer questions in the public interest, as hurdles to achieving ecological validity in research settings thwart the validity of the findings themselves. I have chosen to conduct a live experiment about spear phishing and disinformation on Twitter *in real time*, *on a real setting*, and *during the very real political climate* leading up to the 2018 Midterm Elections. Certainly this provokes ethical risks, but careful attention to a study's design can mitigate risk while producing findings that align closer to real-world conditions than controlled experimental settings. Careful attention to a study's design can help tip this risk/reward ratio in favor of the latter, such as: attempts to maintain political neutrality, using only training corpora generated by users themselves, and careful filtering of users to minimize the impact of treatment.

In conclusion, one of the primary motivations of this study is to lobby social science researchers to help take up the fight against cyber threats and disinformation. In an era of API lockdowns, scholars need to move beyond platform reliance on data in future research and actively conduct experiments to answer questions in the public interest. This study marks an exploratory step toward doing so, and it is limited by several factors. The study examines one platform and one national context, of which participant population is not representative. Further, the experimental design encountered resistance by Twitter itself and was blocked from completion. Still, the study's findings call for more research to uncover the mechanisms that underpin users' susceptibility to spear phishing, disinformation, and online news seeking more broadly.

*About the author:*
Michael Bossetta is a PhD Fellow at the Department of Political Science, University of Copenhagen. His research interests primarily revolve around citizens' use of social media during elections, the design of social media platforms, and how platform design is manipulated to conduct cyberattacks. He is the producer and host of the Social Media and Politics podcast. You can follow him on Twitter @MichaelBossetta and the podcast @SMandPPodcast.

*Endnotes*

[1] Facebook, 2018, https://newsroom.fb.com/news/2018/07/removing-bad-actors-on-facebook/
[2] Harvey and Roth, 2018, https://blog.twitter.com/official/en_us/topics/company/2018/an-update-on-our-elections-integrity-work.html
[3] Verizon, 2018, p. 12
[4] Social Science One, 2018, https://socialscience.one/our-facebook-partnership
[5] Gadde and Gasca, 2018, https://blog.twitter.com/official/en_us/topics/company/2018/measuring_healthy_conversation.html
[6] Bossetta, 2018, p. 473
[7] Seymour and Tulley, 2016, p. 1
[8] Varol *et al.,* 2017, p. 1
[9] Chhabra *et al.,*, 2011, p. 97
[10] Bruns and Moe, 2014, p. 19
[11] Seymour and Tulley, 2016, https://github.com/zerofox-oss/SNAP_R
[12] Google, 2018, https://firebase.google.com/docs/dynamic-links/
[13] Hootsuite, 2018, https://blog.hootsuite.com/best-time-to-post-on-facebook-twitter-instagram/#twitter
[14] Twitter, 2018, https://help.twitter.com/en/managing-your-account/using-the-tweet-activity-dashboard

23